# Numerical and Fitting Approach for Investigation of the Schrödinger Equation for Flat Potentials

## Vahid MIRZAEI MAHMOUD ABADI[*]

## Javad MOHAMMADI

*Faculty of Physics, Shahid Bahonar University of Kerman, P.O. Box 76175, Kerman, Iran*

E-mail:vah_mirzaei@uk.ac.ir

**Abstract**

In this study, the Schrödinger equation for flat potentials through the pseudospectral method is investigated. The Schrödinger equation for this type of potentials by the pseudospectral method is transformed to an eigensystem. Then the eigensystem using the Jacobi method is diagonalized. Energy eigenvalues for various *N* are compared with similar researches. Asymptotic treatment of this type of potentials for *N=2* and *N→∞* which correspond to harmonic oscillator and particle in the infinite square well is discussed. For every *N* an equation for quantized energies versus quantum number *n* is suggested. Using fitting the calculated data for the energy eigenvalues accuracy criterion of the suggested energy equation is checked.

## Introduction

Finding the wave functions and energy eigenvalues for every potential is one of the most important problems in quantum mechanics. Some potentials such as the harmonic oscillator and the infinite square well and Coulomb are very important because of its applications [1-3]. The Schrödinger equation has the analytical solution for a few potentials, but the Schrödinger equation for almost potentials does not have the analytical solution. Therefore in these cases perturbation, approximation, and numerical methods are used. In the past decades, much effort has been made to solve this equation [4-6]. Rresearches on numerical solution of equations are performing [7,8]. Radial potentials are among the well-known potentials that are widely used in various physics trends including nuclear, molecular atomic, solid state, and etc. these group of potentials is applied to discover many phenomena [9]. Often these potentials do not have an analytical solution.

The pseudospectral method is a numerical method which is used for solving the Schrödinger equation and has high accuracy. This method is based on the using of orthonormal functions to establish an eigensystem as an estimate of the Schrödinger equation. The accuracy of each orthonormal function sets is depended on the defined domain of the spatial coordinate. For radial potentials in spherical coordinates such as Coulomb potential in hydrogen atom which coordinate *r* belongs to (0, ∞), the Laguerre polynomials are suitable. In harmonic oscillator problem, spatial coordinate belongs to (-∞, ∞), in such cases, Hermit polynomials are suitable orthonormal set, and so on. In this study, the sine functions are used as the orthonormal set. One can find the detailed explanation in [10].

The potential case study is the batch of potentials with the general form equation (1) [11, 12].

$$V(x) = \mu \left|\frac{x}{a}\right|^N, \quad N = 2, 3, \ldots \qquad (1)$$

where $\mu$ and $a$ are constant parameters and $N$ is an integer number larger than one. This batch of potentials has applications in several cases [1-3, 13]. Some of the general application of this type of potentials are structural phase transitions [14] and polaron formation in solids [15].

For *N=2* which is harmonic oscillator the energy eigenvalues are obtained analytically and are as follow [16]

$$E_n = \hbar a\sqrt{\frac{2\mu}{m}}\left(n+\frac{1}{2}\right), \quad n = 0, 1, 2, ... \qquad (2)$$

For *N=4* the potential is named quartic anharmonic potential [12, 17-19]. For *N=6* the potential is called sextic potential. This potential is applied in fiber optic [20] and molecular physics [21]. The case *N=8* is octic potential that recently has been become an interesting subject in Lower-dimensional field theory [22].

Potential equation (1) has two extremes, *N=2* which correspond to the harmonic oscillator, and *N*→∞ which corresponds to infinite square well with *2a* width [2, 3]. The Schrödinger equation can be solved analytically at two these extremes. But, between these two extremes for *N=3, 4...* there is no analytical solution of the Schrödinger equation. The linear combinations of the various power of *x* in the potential of equation (1) also it has become important in several issues.

Increasing *N* in equation (1) shows the energy eigenvalues converge to the ground state energy eigenvalue of the infinite square well. In the present study, some of lower power of *x* in equation (1) (*N=2, 3... 11*) is numerically investigated. The Schrödinger equation has been solved numerically through the pseudospectral method. The calculated data related to the energy eigenvalues result of the numerical solution of the Schrödinger equation has been fitted to find an equation for quantized energies.

## Testing the numerical results

The accuracy of the numerical solution is performed by comparing the numerical results with the analytical results of the case of the harmonic oscillator. For *N=2* with $\hbar=m=1$ Schrödinger equation and the analytical energy eigenvalues equation are respectively as follow

$$-\frac{1}{2}\frac{d^2U(x)}{dx^2} + x^2 U(x) = EU(x) \qquad (3)$$

$$E_n = \sqrt{2}\left(n+\frac{1}{2}\right), \quad n = 0,1,2, ... \qquad (4)$$

where $\mu=a=1$. Some of the lowest exact energy eigenvalues of the harmonic oscillator potential are presented in table 1. In table 2 the numerical results for *N=2, 3...11* and the lowest eight energy eigenvalues are given. The numerical calculation has been performed by assumption $\mu=a=\hbar=m=1$. As it seen that the exact energy eigenvalues in table 1 are exactly the same as numerical results in the second column of table 2. Numerical results have been obtained with 1000 mesh points. Energy eigenvalues in table 1 and 2 are given with twelve decimal digits.

For some *N>2* the ground state energy is reported by authors. In reference [23] energy of the ground state with high accuracy has been numerically calculated and reported for *N=4*

$E_{g.s}$= 0.667986259155777.....

The results of our calculation as it given in table 2 are exactly the same as reference [23] until twelve decimal digit. It should be noted that the accuracy of the calculated energy eigenvalues with getting away from ground state decreases.

**Table 1.** Exact energy eigenvalues of the harmonic oscillator for $\mu=a=\hbar=m=1$.

| $n$ | $E$ |
|---|---|
| 0 | $\frac{1}{2}\sqrt{2} = 0.707106781186$ |
| 1 | $\frac{3}{2}\sqrt{2} = 2.121320343559$ |
| 2 | $\frac{5}{2}\sqrt{2} = 3.535533905932$ |
| 3 | $\frac{7}{2}\sqrt{2} = 4.949747468305$ |
| 4 | $\frac{9}{2}\sqrt{2} = 6.363910306789$ |
| 5 | $\frac{11}{2}\sqrt{2} = 7.778174593052$ |
| 6 | $\frac{13}{2}\sqrt{2} = 9.192388155425$ |
| 7 | $\frac{15}{2}\sqrt{2} = 10.60660171780$ |

The treatment of energy eigenvalues results of variation of two parameters $a$ and $\mu$ is investigated. The curves in figure 1 are the 3-D plot of ground state energy versus parameters $a$ and $\mu$.

**Table 2.** First eight energy levels of potential of equation (1) for $N=2, 3...11$.

| | $N=2$ | $N=3$ | $N=4$ | $N=5$ | $N=6$ |
|---|---|---|---|---|---|
| $E_0$ | 0.707106781186 | 0.674893907795 | 0. 667986259155 | 0.671857829758 | 0.680703611664 |
| $E_1$ | 2.121320343559 | 2.276522383002 | 2.393644016483 | 2.492370519537 | 2.579746228437 |
| $E_2$ | 3.535533905932 | 4.202826147289 | 4.696795386863 | 5.081870058717 | 5.394888357434 |
| $E_3$ | 4.949747468306 | 6.282227611687 | 7.335729995227 | 8.183904853398 | 8.880504996803 |
| $E_4$ | 6.363961030679 | 8.491229118714 | 10.24430845544 | 11.69612259804 | 12.91132000821 |
| $E_5$ | 7.778174593051 | 10.79975828964 | 13.37933655260 | 15.56420364876 | 17.42167370791 |
| $E_6$ | 9.192388155424 | 13.19565902369 | 16.71188963292 | 19.74822516682 | 22.36487507771 |
| $E_7$ | 10.60660171780 | 15.66616870313 | 20.22084946407 | 24.21917775109 | 27.70567848983 |

Continue of table 2

| | $N=7$ | $N=8$ | $N=9$ | $N=10$ | $N=11$ |
|---|---|---|---|---|---|
| $E_0$ | 0.691873845711 | 0.704048774123 | 0.716533639005 | 0.728951380349 | 0.741095970320 |
| $E_1$ | 2.658906121341 | 2.731532557944 | 2.798679010580 | 2.861086465499 | 2.919319767619 |
| $E_2$ | 5.657775904672 | 5.884176869892 | 6.082964051387 | 6.260149428344 | 6.419961157344 |
| $E_3$ | 9.463809070926 | 9.960988311189 | 10.39147614372 | 10.76935524114 | 11.10499267188 |
| $E_4$ | 13.94076667278 | 14.82338179966 | 15.58880571133 | 16.25968621504 | 16.85344497407 |
| $E_5$ | 19.01295392472 | 20.38818898167 | 21.58722779580 | 22.64152738136 | 23.57597038778 |
| $E_6$ | 24.62875535403 | 26.59970164150 | 28.32763019106 | 29.85317224508 | 31.20917724931 |
| $E_7$ | 30.74861276198 | 33.41543409575 | 35.76542100309 | 37.84841743815 | 39.70562049989 |

Each curve includes 2500 point which is fifty step on the parameter $a$ and fifty step on the parameter $\mu$. Every point of 3-D curves has been obtained of 200 mesh. It is observed that more little values of parameter $a$ correspond to the greater ground state energy and vice versa. For the constant value of $a=0.1$, with increasing parameter $\mu$, the ground state energy also increases. In the constant value of $a=0.1$, for $N=2$, with increasing parameter $\mu$ from *0.1* to *5*, the ground state energy increases from *3.137* until *7.906*. For $N=3$ at the same values of the parameter $\mu$, the ground state energy varies from *4.258* until *20.370*. For the $N=4$ variations of the ground state energy is from *6.680* to *24.609*. In the case $N=5$ it is seen that the ground state energy increases from *9.335* until *28.542*. For $N=6$ at the same situation, the ground state energy varies from *12.105* until *32.150* and finally, for $N=7$ the ground state energy increases from *14.907* until *35.385*. Therefore at the constant value of parameter $a=0.1$ general treatment of the ground state energy versus the parameter $\mu$ is ascending. With increasing $N$ the variations domain of the ground state energy is also ascending.

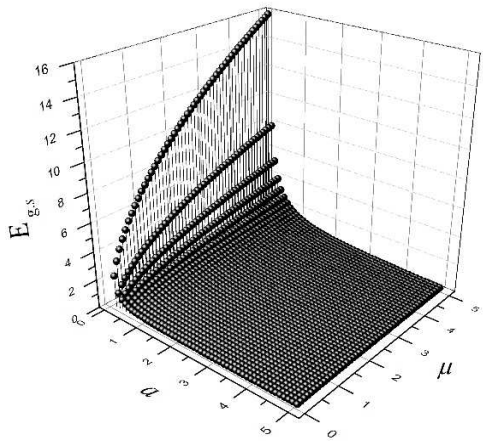

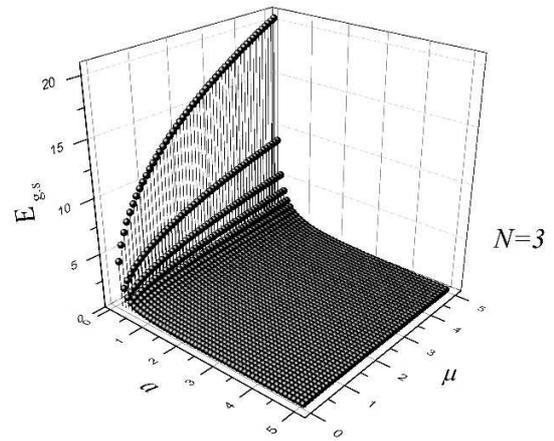

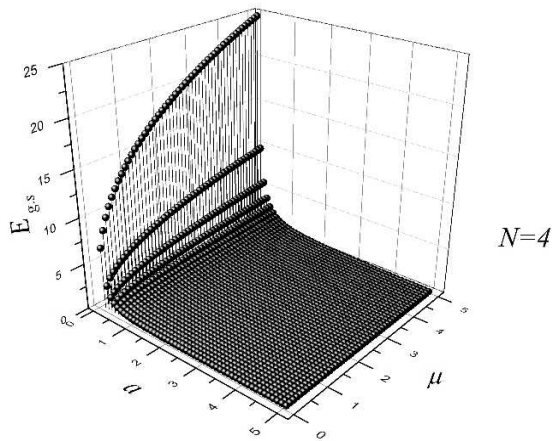

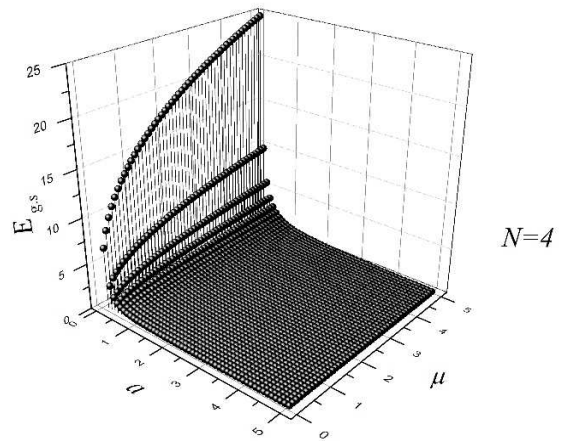

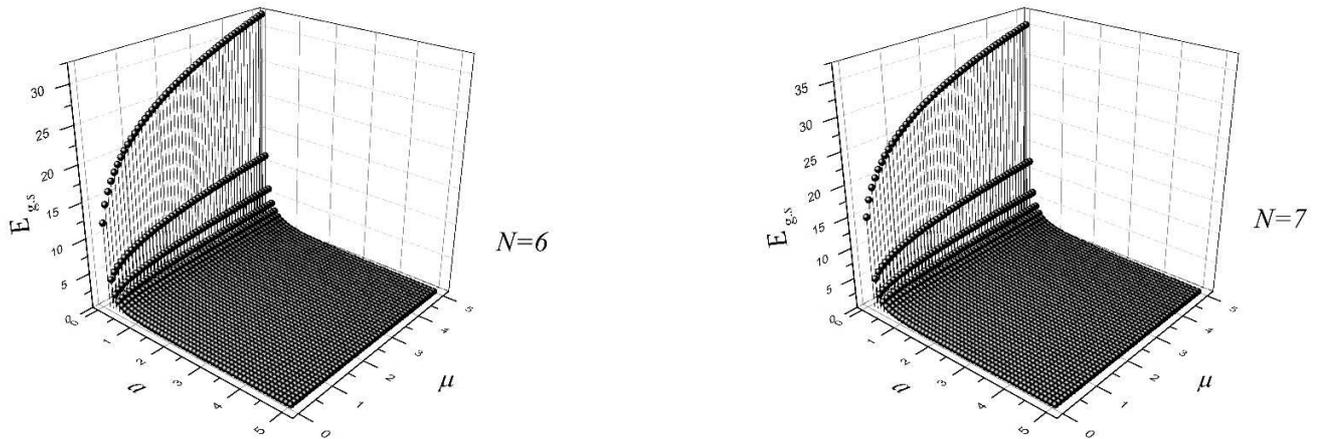

**Figure 1.** Ground state energy versus parameters $a$ and $\mu$ for $N=2, 3...7$ in equation (1).

In figure 2 the potential equation (1) for some $N$ has been shown.

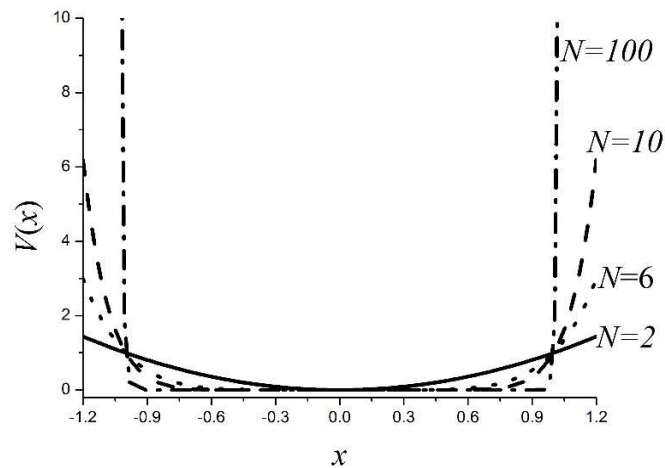

**Figure 2.** Potential of equation (1) for some values of $N$.

With increasing $N$ in figure 2, the curve of the potential of equation (1) as is expected tend to infinite square well. The shape of the wave function with increasing $N$ reaches to the wave function of the infinite square well. In figure 3 the ground state normalized wave function for some $N$ is shown. From little $N$ to the large one it is seen that spatial expansion of the wave function step by step gets less and the peak of the wave function gets sharper.

If the energy levels are plotted for each $N$, depending on the value of $N$ at the potential of equation (1), the distance between the consecutive levels will be different. For $N=2$ which corresponds to the harmonic oscillator, the difference of consecutive energy levels is constant.

The increase in successive energy levels is dependent on $N$. In figure 4 the graph of first energy levels for $N=2, 3...$ $11$ has drawn up. For each $N$, the number of eight primary energy levels is calculated and given. Increasing the distance between the energy levels by increasing $N$ is clearly seen in figure 4. The horizontal axis is the parameter $N$ and the vertical axis is energy in figure 4. Quantized energies eigenvalues versus quantum number $n$ as the curve is shown in figure 5 for $N=2, 3…11$. The linear dependence of energy versus quantum number $n$ for $N=2$ at the lowest of the curve is seen. As $N$ increases, the energy versus the quantum number $n$ is exited from the linear dependence. At the other extreme, when $N$ tends to infinity (the infinite square well potential), we know that the energy in terms of the quantum number $n$ is a second order.

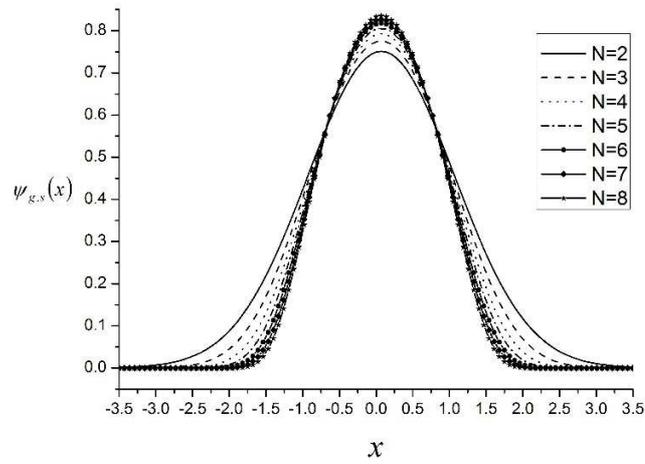

**Figure 3.** Normalized ground state wave function of the potential of equation (1) for some *N*.

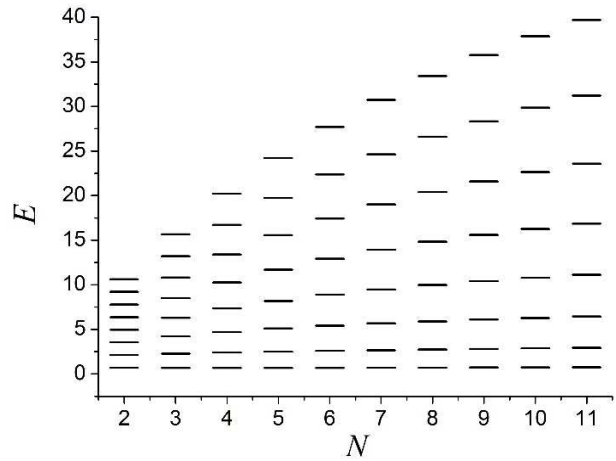

**Figure 4.** Some first energy levels of the potential of equation (1) for *N=2, 3…11*.

According to figure 5, it seems that in interstitial values of *N* in equation (1) between *N=2* and *N→∞* depending on energy eigenvalues in terms of quantum number *n* must vary from *1* for the harmonic oscillator (*N=2*) to *2* for the infinite square well (*N→∞*). Therefore, it is suggested that the energy eigenvalues in terms of the quantum number *n* have a form as follow

$$E_n = \alpha\left(n^{\beta(N)} + \gamma\right) \qquad (5)$$

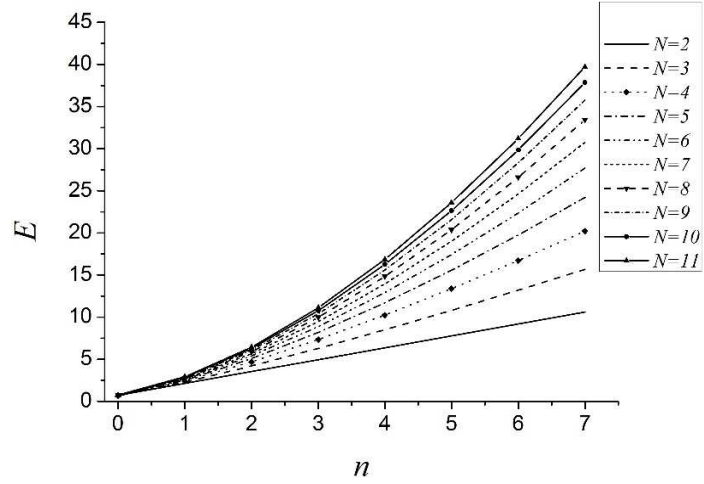

**Figure 5.** Curves of energy versus the quantum number *n* for *N=2, 3…11*.

In the case of the harmonic oscillator (*N=2*) for *a=μ=ℏ=m=1*, the parameters *α*, *β*, and *γ* are as follow

$$\alpha = \sqrt{2}, \quad \beta = 1, \quad \gamma = \frac{1}{2} \qquad (6)$$

On another extreme for *N→∞* and for *a=μ=ℏ=m=1* we have

$$\alpha = \frac{\pi^2}{8}, \quad \beta = 2, \quad \gamma = 0 \qquad (7)$$

The energy eigenvalues which are calculated numerically through pseudospectral mothed have been fitted base on suggested equation (5). The results of fitting energy eigenvalues versus the quantum number *n* for each *N* are given in table 3. In table 3 the statistical criteria for the accuracy of fitting are given for some first *N*'s. According to the performed fitting, the predicted treatment of energy eigenvalues versus the quantum number *n* which suggested in equation (5) can be seen. In other words, the power of the quantum number *n* (*β* in equation (5)) of the harmonic oscillator (*N=2*) from the value *1* gradually increases and asymptotically tends to *2* for the infinite square well.

**Table 3.** The equation of the energy eigenvalues in terms of the quantum number n for *N=2, 3…11* in equation (1).

| Power in potential | Energy levels equation | Adj. R-Square | Reduced Chi-Sqr. |
|---|---|---|---|
| *N=2* | $E_n = 1.41421(x^{1.00000} + 0.50000)$ | *1.00000* | *0.00000* |
| *N=3* | $E_n = 1.56301(x^{1.16072} + 0.44281)$ | *0.99999* | *0.00034* |
| *N=4* | $E_n = 1.63934(x^{1.27203} + 0.43185)$ | *0.99997* | *0.0014* |
| *N=5* | $E_n = 1.68442(x^{1.35312} + 0.43722)$ | *0.99996* | *0.00316* |
| *N=6* | $E_n = 1.71425(x^{1.41452} + 0.44936)$ | *0.99994* | *0.00551* |
| *N=7* | $E_n = 1.73577(x^{1.46244} + 0.46440)$ | *0.99993* | *0.00838* |
| *N=8* | $E_n = 1.75257(x^{1.50071} + 0.48051)$ | *0.99991* | *0.01169* |
| *N=9* | $E_n = 1.76659(x^{1.53185} + 0.49672)$ | *0.99990* | *0.01537* |
| *N=10* | $E_n = 1.77892(x^{1.55755} + 0.51252)$ | *0.99989* | *0.01935* |
| N=11 | $E_n = 1.79025(x^{1.57901} + 0.52760)$ | 0.99988 | 0.02354 |

In figure 6, the power of the quantum number *n* in the eigenvalue energy equation which is given in table 3 for *N=2, 3...11* versus *N* in equation (1) are shown. The best fitting for the power of the quantum number *n* versus *N* also is given in figure 6. Fitting accuracy criteria are also given in figure 6.

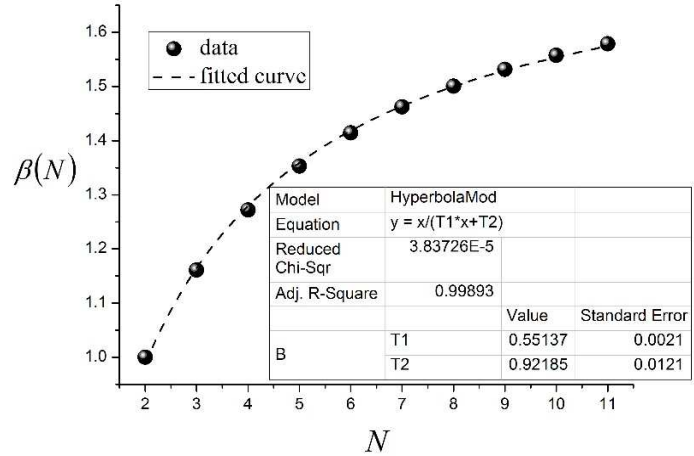

**Figure 6.** Fitted curve and data curve of *β*, the power of the quantum number *n* versus *N* in equation (1).

Regarding the endpoints of *N* namely *N=2* and *N→∞*, the relation between the parameter *β* and *N* in equation (1) seems to be

$$\beta(N) = \frac{N}{0.5N+1} \qquad (8)$$

**Summary and conclusion**

In this study, Schrödinger's equation for flat potentials in the form of equation (1) was numerically solved through the pseudospectral method. Comparing the obtained results of this method with the analytical solution for *N=2* and also similar numerically researches for *N>2* confirm the high accuracy of this method. The calculated data from the pseudospectral method presented in table 2 shows a good agreement for *N=2*, which correspond to a simple harmonic oscillator. On the other hand, in certain cases, some of the energies of this potential, especially for the ground state, has been reported in some papers for *N≠2* and the results of the present study have been compared with similar works which are in good agreements.

The proposed equation (5) which is presented for the first time, gives the relation between energy eigenvalues and the quantum number *n*. This suggestion is based on the form of the analytical equation of the energy eigenvalue for the two endpoints of the parameter *N* in equation (1), namely *N=2* and *N→∞*. For *N=2* energy eigenvalue is obtained analytically and the relation is linear. On the other hand, as *N→∞* we are faced with the infinite square well. The relation between energy eigenvalue and the quantum number *n* is in the form of second-order $E=cn^2$. Fitting on the data presented in Table 3, according to statistical criteria due to fitting confirms the suggested equation (5).